# A Review of Blockchain-based Smart Grid: Applications, Opportunities, and Future Directions

A PREPRINT

**H. Sami Ullah**
Dept. of Computer Science
University of Gujrat, Gujrat, Pakistan
hamzasamiullah@gmail.com

**S. Aslam**
Dept. of Computer Science
University of Gujrat, Gujrat, Pakistan
samiaaslam479@gmail.com

## Abstract:
The Smart Grid (SG) concept presented an unprecedented opportunity to move the energy sector to more availability, reliability, and efficiency to improve our economic and environmental conditions. Renewable energy sources (Solar & Wind) are such technologies that are used in the smart grid to figure out the environmental and economic issues and challenges. Smart grids provide energy in different crowded sectors with efficient and timely transmission of electricity. But the traditional power grids follow a centralized approach for energy transactions with a large number of growing connections and become more challenging to handle power disturbance in the grid. Blockchain as a decentralized and distributed technology provides promising applications in the smart grid infrastructure with its excellent and salient features. In this paper, we provide a concise review of blockchain architecture, concepts, and applications in smart grids. Different potential opportunities for blockchain technology with smart grids are also discussed. Some future directions concluded the paper.

## 1. Introduction

The need for electricity is increasing day by day due to the increasing demand for energy in populated residential and industrial areas [1]. With the advent of contemporary and technical developments, renewable energy also provides an efficient energy generation and consumption mechanism to meet the need for energy by reducing environmental pollution and energy crisis. Renewable energy sources are clean energy sources and facilitate with innovative and less-expensive sources of energy using natural sources. These natural renewable sources such as solar and wind energy will definitely contribute to the current and future energy generation and consumption needs. However, these renewable energy sources with long-distance energy transmission poses some new challenges in the distribution environment of the system [2]. Another challenge is the irregular nature of natural renewable resources such as solar and wind power. The primary task of an energy grid is to provide energy in a stable and efficient manner to the consumers [3]. Smart Grids (SG) provides a contemporary solution to maintain these distributed energy sources and supply energy with enhanced security of the electricity grid in long-distant areas more efficiently. The main goal of a smart grid is to provide local production and consumption of energy by producers to consumers [4]. Transmission losses can be reduced by motivating local energy production and consumption. In the energy trading process, prosumers and consumers should be able to trade the local energy in peer-to-peer fashion without any distortion. A smart grid with centralized infrastructure for transaction of energy between prosumers and consumers requires complex and costly communication infrastructure [5]. Decentralized technologies are preferred to adapt to meet these challenges [6]. Blockchain technology with decentralization mechanisms can provide a solution to the smart grid at different levels of complexity. Blockchain technology has revolutionized and contributed by participating in different fields to enhance commercial and industrial developments. Blockchain development and emergence of utilization in past years is briefly shown in Figure 1:

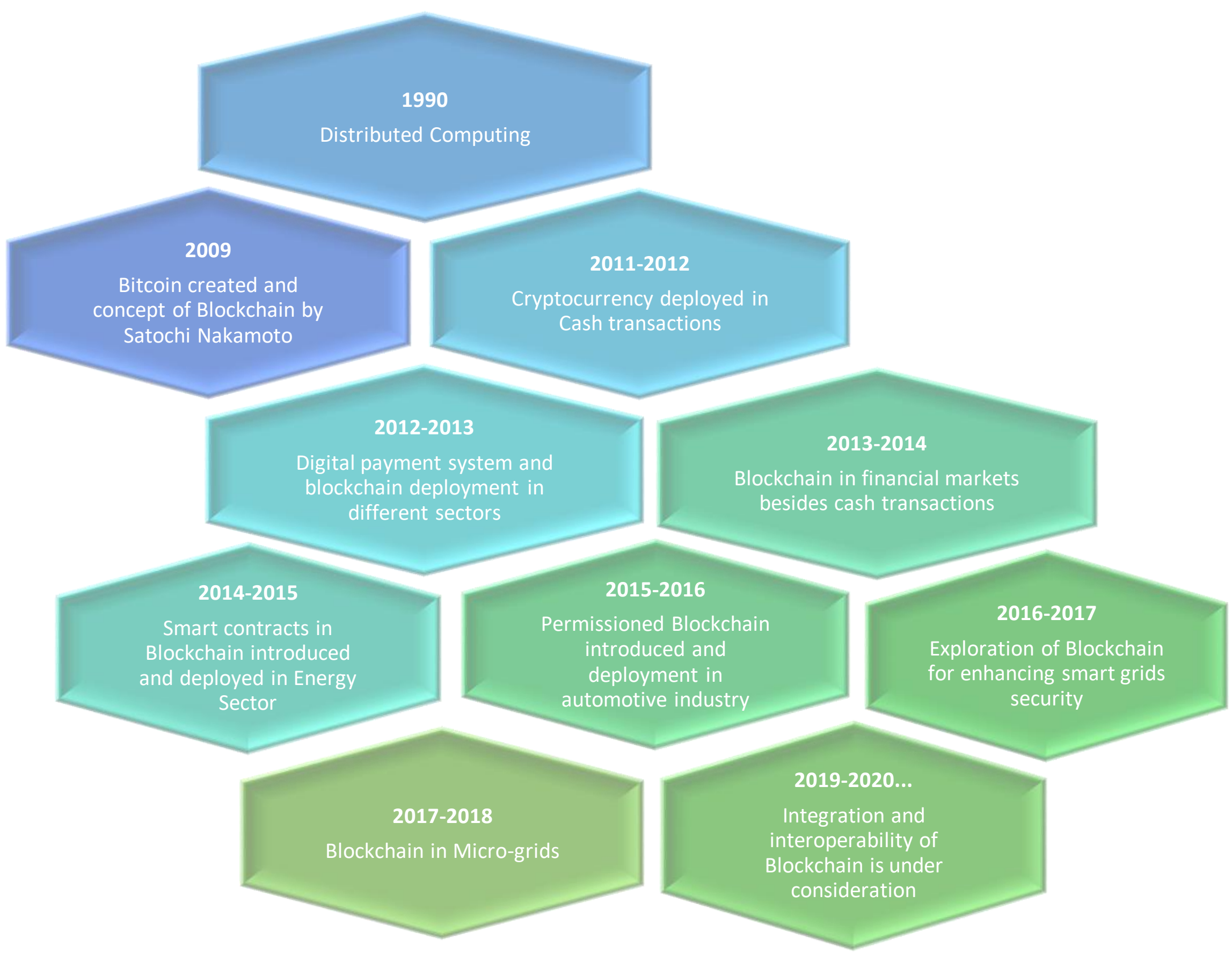

*Figure 1: Evolution of Blockchain*

The original idea of blockchain is the outcome of the concept of distributed computing introduced in the 1990s. Bitcoin a cryptocurrency was proposed by Satoshi Nakamoto [7] in his white paper he introduced blockchain technology for the first time. This paper provided the basis for blockchain evolution and after adopting bitcoin cryptocurrency in transaction application in 2011-2012, the actual application of blockchain in digital payment systems started in 2013 [8]. Smart Contracts (SC) in blockchain have a great impact on many real-life applications such as industrial markets, the Internet of Things (IoT), the energy sector and introduced in 2015. The automotive industry has been revolutionized with blockchain technology such as automotive electric vehicles in 2016 and acquired as a security solution in smart grids in 2017. Micro-grids as a discrete energy network system adopted blockchain in 2018. In 2019, most of the research is done on the integration and interoperability of blockchain in the energy sector and still work is not enough mature to consider to be finished.

Blockchain offers a stable solution in the complex infrastructure of smart grids with its enhanced and secure features and functionalities [9]. Through blockchain integration in smart grids, Prosumers and consumers are able to trade in a peer-to-peer fashion without any centralized authority. The applications of blockchain technology in smart grids provide potential benefits [10]. A smart grid with blockchain integration provides a secure trading infrastructure with limited prosumers and consumers' involvement for effective transactions [11].

The main contribution of this paper is to highlight and summarize the applications of blockchain technology with its architecture, concepts, and opportunities in smart grids. Our paper aims to identify and narrow down different future directions of blockchain technology in the smart grids through the following sequence:

- A brief introduction of blockchain technology with basic concepts, architecture ending with some prominent application areas.
- An in-depth review of applications of blockchain technology in smart grids with different sub-areas.
- Some future directions and conclusion to conclude the paper.

## 2. Blockchain Technology

Blockchain is a ledger-based on distributed database in the shape of connected blocks forming a chain. Each block in the connected chain consists of information and a linked part with the previous and next blocks in the chain [7]. The main advantage of this technology is that it maintains all the variations in the blocks as all blocks are connected in the chain and no block can be modified or removed. This attribute makes blockchain immutable and more secure while transferring safety-critical items such as money and contracts without any third party's intrusion like banks [12].

## 3. Architecture of Blockchain

Blockchain is being a distributed ledger that provides transparency, trust and data security in all applications specifically in safety-critical systems. The Blockchain technology allows the data and information in distributed fashion rather than copied material [13]. Traditional architecture follows client-server architecture in which all the information and data are stored in a single location and that information is easy to modify as the server has the authority of administration with special permissions. But in blockchain architecture, all participating nodes have equal rights to maintain and approve new entries in the chain [14]. The system data is copied to all nodes and no data can be altered or removed once created that makes the blockchain immutable and safe to critical hackers. When a new block comes for entering in the chain, it is verified by all existing nodes with validity and security consensus algorithms and then published in the existing chain. All data in the blockchain is decentralized arranged in a peer-to-peer fashion and the network consists of many computers called nodes in the network [15].

The structure of blockchain technology follows an arranged list of connected blocks with transactions as shown in Figure 2:

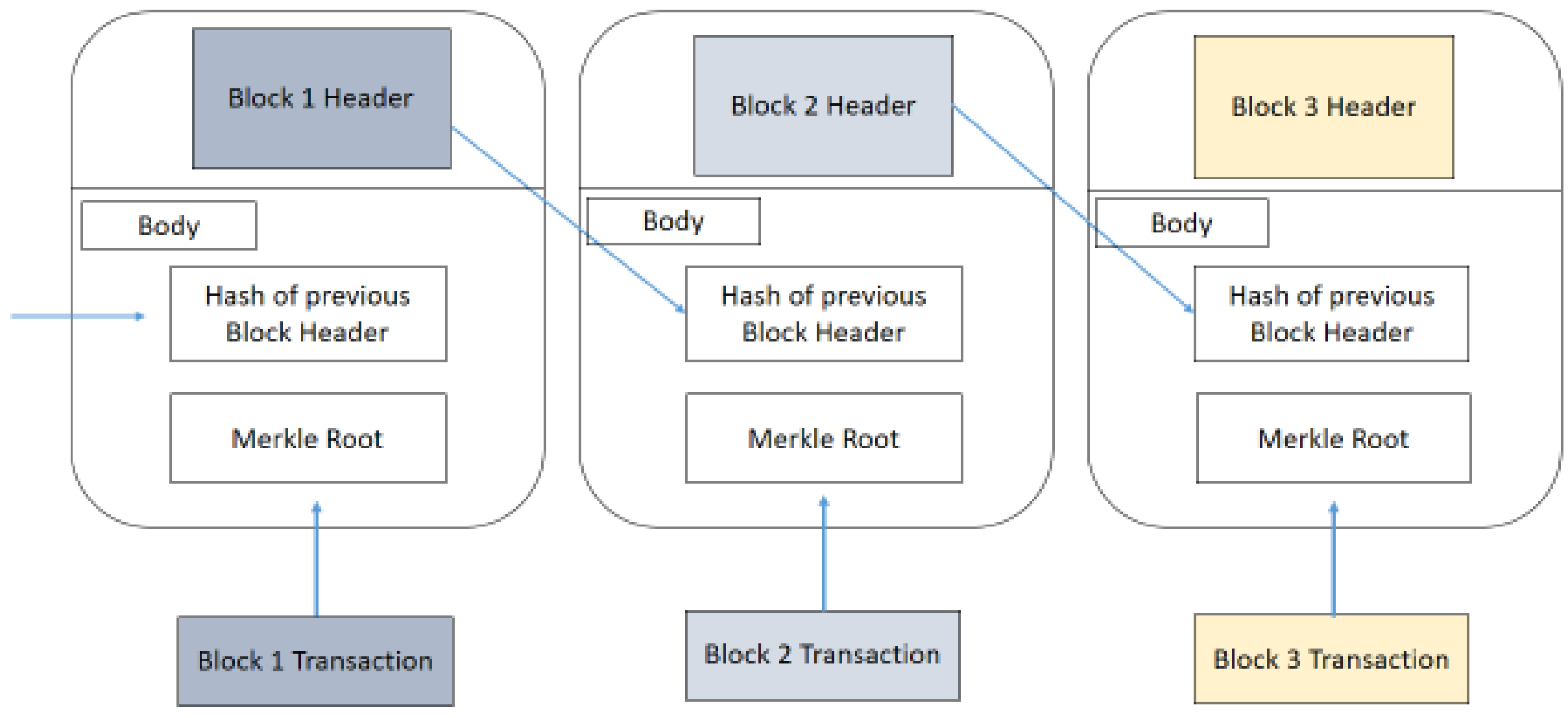

*Figure 2: Basic Structure of Blocks*

Each block in the network is logically divided into two parts, header, and the body part, transactions are stored in the body of the block and the header of each block contains the identifier of the previous block. Blocks are stored in the chain in a sequence of a linked list and the very first block is the genesis block of the chain. For immutability contents, the identifier is obtained by the cryptographic hash value of the block which is linked with all previous blocks [16]. Through this immutability content, a hacker cannot alter the data and information in the single block, he would have to alter all the successive blocks of information as they all are connected in the chain.

## 4. Smart Grid

The Smart Grid (SG) is the modernized concept of the traditional electricity system. A smart grid provides capabilities of two-way communication of electricity data instead of a one-way flow of transactions. Smart grids collect real-time data according to supply and demand during the production and distribution process with efficient electricity generation, consumption, and maintenance. Smart Grid technology is developed to facilitate the prosumers and consumers with efficient, economical and reliable electricity services using available smart tools. Various information about electricity can be gathered with intelligent sensors and a fast distributed communication system for balancing demand and supply of the electricity [17]. The potential demand for the smart grid network has become higher than before due to overloaded power grids which are difficult to handle in the traditional way as this infrastructure is not compatible with current energy needs and requirements [18]. Demand for energy in peak hours generally exceeds the supply from power grids. Different energy generation, distribution and consumption problems may occur at different levels [19]. With the use of fossil fuels for electricity generation in traditional power, grids create environmental and health issues [20]. A smart grid solution provides environmental and economic benefits with effective costs [21]. A smart grid enhances the use of natural renewable energy sources and provides the solution to the problems caused by traditional electric power plants [22], [23]. Renewable energy sources and smart grid utilization in different sectors is shown in Figure 3:

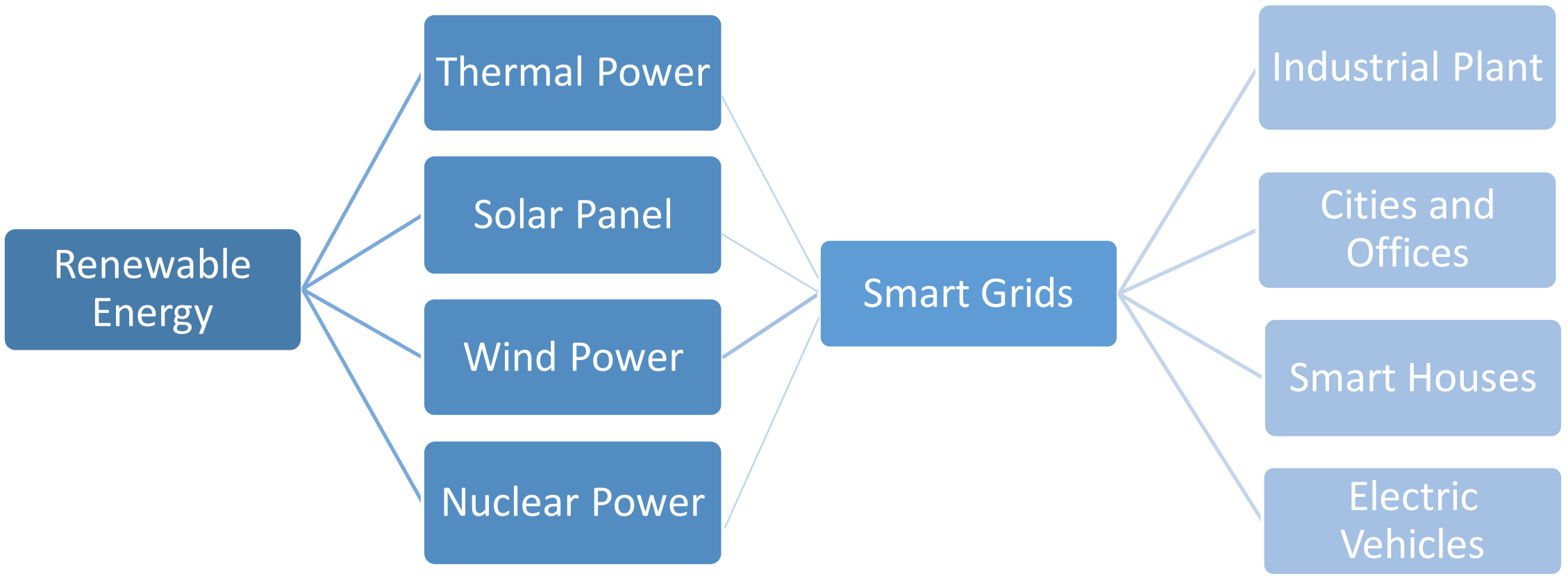

*Figure 3: Smart Grid use cases*

## 5. Blockchain Applications Areas in Smart Grids

Many researchers believe that the development of blockchain technology with its promising features will smooth the shift toward the energy trading in smart grids. Decentralized approaches also provide a basis for smart grid evolution [24]. Electric grid with renewable distributed sources and electric smart vehicles has become a broad research area. Blockchain technology provides a desirable solution to smart grids which captured the interest of its adaptation. Blockchain performs in the smart grid environment in different parts of the grid such as power generation, power transmission and distribution, and power consumption. Blockchain applications are extensively used in the following areas of smart grids. As the smart grid infrastructure with blockchain integration is shown in Figure 4:

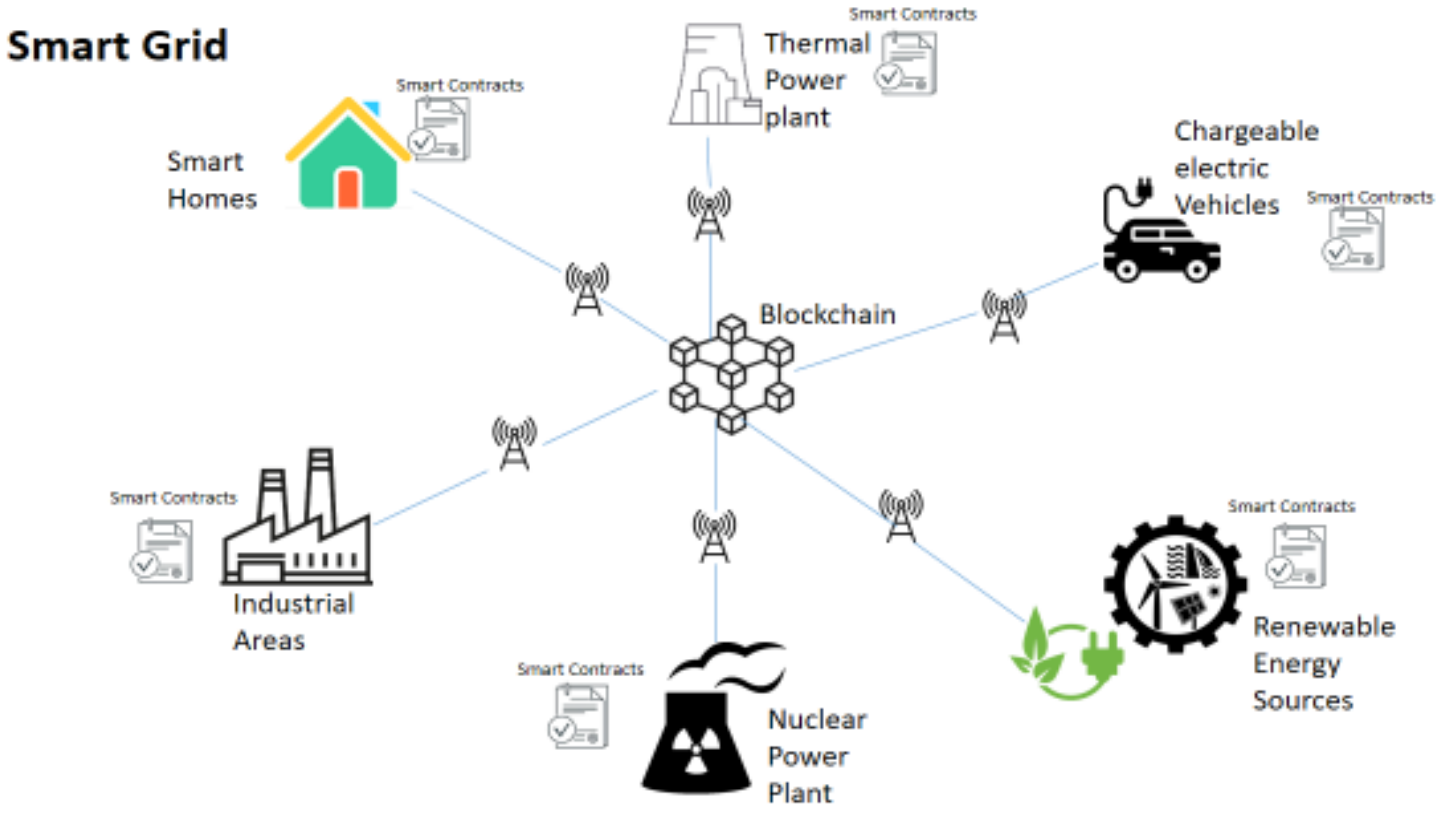

*Figure 4: Blockchain in Smart Grids*

### 5.1. Energy Trading

Energy trading has gained more importance with the distributed energy generation from natural sources and the number of micro-grids for trading energy for domestic and industrial consumption [25]. Blockchain has a great impact on energy trading for eradicating false transaction behaviors of energy trading [26]. A Proof of Origin (PoO) certificate for assurance of fair trading process in the energy market is issued [48]. Blockchain in the energy trading process eliminates the third parties and P2P connections allow producers and consumers to make deals with other nodes in the chain independently with freedom of their preferences.

### 5.2. Electric Vehicles

Electric vehicles are also a connected part with smart grids and gained more attention in the contemporary research [28]. The charging of the electric vehicles is the key concern when it comes to the smart grid which increases the consumption of energy from the main power grid [29]. The integration of blockchain is proposed for vehicles to reduce power fluctuations by finding their nearest available charging station through bidding for charging [30]. Vehicles can find a more suitable location and price for their charging with security and privacy concerns [31].

### 5.3. Micro-grid Operations

Micro-grids are supposed to provide the energy produced by distributed energy sources. Therefore the control and operations of micro-grid based on users' demands is the hot topic today for research [32]. Blockchain provides a secure and trustworthy scheduling mechanism for distributed energy sources to be trusted and secured. Blockchain provides the benefit of reducing the ratio of average-to-peak in the grid and smooth the power fluctuations caused by constraints of supply [33].

### 5.4. Cyber-Physical Infrastructure

With the integration of smart grids, many impurities and vulnerabilities may occur at the cyber-physical layer of the smart grid and can manipulate different parts of the grid through attacks. These attacks are considered in detail in [34]. Cyber-physical attacks are different in nature, and the impact such as GPS spoofing attacks [35], Denial-of-Service (DOS) attacks [36], time synchronization attacks [37]. Blockchain-based infrastructure of smart grids are utilized to overcome any of these attacks in the smart grid without the interruption and manipulation from any outside party [38]. Blockchain with smart contracts is more beneficial to provide a resilient network for all energy and information transactions performed in the grid [39].

*Table 1: Different applications of Blockchain in Smart Grid*

| Ref | Blockchain with Smart Grid |
|---|---|
| [40] | Ethereum private blockchain for trading local energy markets with on-chain simulation |
| [19] | Blockchain is used with a consensus mechanism and reward mechanism as a tool for peer-to-peer transactions of energy in smart grid by developing a mobile application for testing |
| [41] | For cyber resilient transactions in smart grids, blockchain with smart contracts to improve security is proposed with PNNL Buildings-to-grid (B2G) testbed tested on Splunk |
| [42] | Proof of Energy Generation (PoEG) and Proof of Energy Consumption (PoEC) algorithms for prosumers blocks validation are proposed using data of Ontario Electric Board |
| [25] | A trading process in smart grids is designed, built and tested with Multichain architecture of blockchain |
| [43] | Solidity platform is proposed and used for implementing smart energy contracts in smart grids |
| [44] | The blockchain concept in smart grid is implemented with TestRPC- Ethereum |

# 6. Opportunities in Blockchain

In this section, some potential and prominent opportunities for adopting blockchain technology in smart grids are presented. We will discuss blockchain with its security and trust features that will ultimately encourage to move the power grid network to blockchain in smart grid infrastructure.

### 6.1.1. Decentralized Smart Grid

As we discussed in the introduction section that the distributed smart grid with digital communication and computation technologies can transform traditional power grids to the more efficient and accurate energy solution. These changes in the energy grid has come due to climate change and the need for sustainable and effective energy sources. The ultimate objective of this modernized smart grids is to utilize more renewable and natural sources for energy production and consumption and making the environment clean. As conventional power grids use long-distance power lines that can be damaged or failed to provide energy to the consumers in time, but the smart grid infrastructure brings prosumers and consumers closer to provide extra saved renewable energy independently without central power grid involvement.

The energy internet (EI) concept is introduced in [45] and can be defined as the advanced version of smart grids. The EI network is a collection of energy, information, and economies to develop an improved version of the smart grid. The main objective of EI is to provide integration between renewable sources and distributed energy sources with the smart grid and to provide more effective interactions among different elements of the grid to make transactions more intelligent and robust in nature. The main principle behind the EI technique is to share information and energy on the internet like other things shared on the internet. These different interacting elements may be energy generation units, distributed energy sources, electric chargeable vehicles, Vehicle to Grid (V2G) environment [46], prosumers, consumers and energy trading markets etc.

Although smart grids and energy internet utilizes more sources as compare due to traditional power grids, but somehow these are dependent on some centralized entities [47]. These centralized systems monitor, collect and process all the information and the energy is transferred to the consumers via long-distance distributed network [48]. By transferring this traditional smart grid to a decentralized network provides a proactive and robust solution by removing central power authorities [49].

### 6.1.2. Motivations for Blockchain in Smart Grids

Security, trust, and privacy of the system are initials to consider for any transaction of information in any system. Decentralized smart grids may face the challenges of unauthorized user access, undesired modification in transactions, evidence issues for some specific transactions, fault-tolerant issues in attacks, lack of efficient monitoring of the network, lack of privacy-preserving techniques, transparency issues among nodes. All these challenges in the smart grid environment can be solved with the consensus-based blockchain technology which uses cryptographic secure data structures, time stamps, digital signatures, and efficient privacy-preserving techniques. Applying blockchain technology as an opportunity that facilitate the transmissions in the smart grids with its following salient and suitable features:

- The blockchain is basically formed with various decentralized nodes and these nodes run in a peer-to-peer manner without trusting any centralized authority for permissions and maintenance.
- The blockchain is scalable in nature such as more nodes can be added easily to enlarge the network according to the new demands in the energy sector and new blocks are maintained and validated by existing peer nodes.
- Blockchain network is trustless [50] in nature but provides better security as compared to a centralized network as all nodes are independent and transactions are secured cryptographically without any central authority.
- In blockchain data nodes created once cannot be altered or removed until most of the nodes become malicious.
- Blockchain provides a resilient and better fault-tolerant network in which an error or malicious attack can be identified and removed easily. The entire chain of blocks is arranged and published with premises with no point of failure at any single node.

## 7. Future Directions

Some recent trends and future directions in blockchain technology and smart grids applications are highlighted. Blockchain interoperability refers to the ability to connect and share data and information across multiple blockchain networks and systems. For example nodes in one blockchain network can interact with another blockchain network via interoperability and a more vast network is can be built for a more secure and distributed network of multiple chains. Optimal placement and sizing of renewable distribution sources can be reconsidered for optimal solutions in the smart grids. Integrated forecasting mechanisms for future energy production and consumption in the smart grid can be applied for better availability of energy in industrial areas.

## 8. Conclusion

With the abundance of blockchain-based applications and experiences in different fields, trust in the longevity of this distributed ledger is increasing. Primary features of security, privacy, immutability and trust in the transactions of information have potential benefits to the energy sector. Through blockchain energy transactions in smart grids are performed in a peer-to-peer fashion without any third trusted authority. These unique and salient features of blockchain provided researchers a contemporary insight into the research area. Energy grids are producing and utilizing energy more sufficiently with the integration of blockchain. In this paper, blockchain architecture, smart grid concepts and evolution of blockchain in a smart grid with different application areas is reviewed. Furthermore, we illustrated some potential benefits and opportunities for the adaptation of blockchain technology in modern power grids. Some prominent future directions of blockchain and smart grids are also highlighted. However, some significant challenges will definitely arise in the coming years when these directions will be tested.